\documentclass[12pt]{article}
\usepackage{graphicx}

\begin{document}
\title{Form factors of pion and kaon}
\author{Jun Gao and Bing An Li\\
Department of Physics and Astronomy, University of Kentucky\\
Lexington, KY 40506, USA}

\maketitle
\begin{abstract}
Besides the vector meson poles in the form factors of charged pion
and kaons additional intrinsic form factors of pion and kaons are
found.  
A detailed study of the form factors of pion
and kaons 
in both time-like and space-like regions 
is presented. Theory agrees well
with data up to $q\sim 1.4GeV$. In this study there is no adjustable parameter.
\end{abstract}

\pagebreak
Form factor is a very important physical quantity in
understanding the internal structure of hadrons. The vector meson
dominance (VMD)[1] is successful in studying electromagnetic
interactions and the  vector part of weak interactions of mesons.
According to the VMD the pion form factor is determined by a
$\rho-$meson pole[2]. Generally speaking, the $\rho-$ pole fits
the data pretty well. However, a detailed study shows that there
is deviation between the experimental data of the pion form factor
and the $\rho-$pole. The electromagnetic radius of charged pion
has been determined to be[3]
\begin{equation}
<r^2>_\pi ^{exp}=(0.439\pm 0.03)fm^2,
\end{equation}
whereas the result obtained from $\rho$-pole is
\begin{equation}
<r^2>_\pi ^{\rho -pole}=0.391fm^2.
\end{equation}
In Ref.[4] a comparison between the measurements and the $\rho-$
pole[2] of the pion form factor is presented. The Fig.6.3 of Ref.[4]
shows that the $\rho-$ pole[2] decreases a little bit faster in
time-like region and slower in space-like region. The experiments
show that besides the $\rho-$ pole maybe an additional
form factor is needed.

Based on t'Hooft's arguments that in large $N_{C}$ limit QCD is
equivalent to a meson theory[5] an effective chiral theory of
large $N_{C}$ QCD of mesons  has been proposed[6]. It has been
shown that the tree diagrams of mesons are at leading orders
and the loop diagrams of mesons are at higher orders in large $N_{C}$
expansion. In chiral limit, $m_{q}\rightarrow 0$, the Lagrangian
of this theory in the case of two flavors takes the form
\begin{eqnarray}
{\cal L}&=&\overline{\psi }(x)[i\gamma \cdot \partial +\gamma
\cdot v+\gamma \cdot a\gamma
_5-mu(x)]\psi (x) \nonumber \\
&&+\frac 12m_0^2(\rho _i^\mu \rho
_{\mu i}+\omega ^\mu \omega _\mu+a_i^\mu a_{\mu
i}+f^\mu f_\mu ),
\end{eqnarray}
where $a_\mu =\tau _ia_\mu ^i+f_\mu ,v_\mu =\tau _i\rho _\mu
^i+\omega _\mu $,
and $u=exp\{i\gamma _5(\tau _i\pi
_i+\eta )\}$; m is a parameter; u can be written as
\[u=\frac 12(1+\gamma _5)U+\frac 12(1-\gamma _5)U^{+},\;\;\;
U=exp\{i(\tau _i\pi _i+\eta )\}.\]
The photon A, W and Z fields can be incorporated to the Lagrangian as well. In
terms of path integral the effective Lagrangian of mesons is obtained by
integrating out the quark fields. All the details can be found in Ref.[6].  The effective Lagrangian has been applied to study meson physics[7,8].
Theoretical results agree well with data.

The VMD is a natural result of this theory[6] and the expressions of the VMD    ($\rho$, $\omega$, and $\phi$) are derived 
\begin{eqnarray}
&&{1\over2}eg\{-\frac 12F^{\mu \nu }(\partial _\mu \rho _\nu
^0-\partial _\nu \rho _\mu ^0)+A^\mu j_\mu ^0\},\nonumber \\
&&{1\over6}eg\{-\frac 12F^{\mu \nu }(\partial _\mu \omega _\nu
-\partial _\nu \omega _\mu )+A^\mu j_\mu ^\omega\},\nonumber \\
&&-{1\over{3\sqrt{2}}}eg \{-\frac 12F^{\mu \nu }(\partial _\mu
\phi _\nu -\partial _\nu \phi _\mu )+A^\mu j_\mu ^\phi\},
\end{eqnarray}
g is a universal coupling constant in this theory and has been
determined to be 0.39 by fitting $\rho\rightarrow ee^+$
($\Gamma^{th}=6.53keV$ and $\Gamma^{ex}=(6.77\pm0.32)keV$). The
current $j_\mu^0$ is derived from the vertex of $\rho\pi\pi$[6]
\begin{equation}
{\cal L}_{\rho \pi \pi }={2\over g}f_{\rho \pi \pi}(q^2)\epsilon _{ijk}\rho
_i^\mu \pi _j\partial _\mu \pi _k
\end{equation}
by substituting
\[\rho^0_\mu \rightarrow{1\over2}egA_\mu\]
into Eq.(5),
where
\begin{equation}
f_{\rho \pi \pi }(q^2)= 1+\frac{q^2}{2\pi ^2f_\pi
^2}[(1-\frac{2c}g)^2-4\pi ^2c^2]= 1+0.243(q^2/GeV^2)  
\end{equation}
where \(c =
\frac{f_\pi ^2}{2gm_\rho ^2}\),
$f_\pi=0.186GeV$, $m_{\rho}=0.773GeV$ and 
$q$ is the momentum of $\rho $ meson. $f_{\rho\pi\pi}(q^2)$ is determined up
to the fourth order in derivatives. It is necessary to emphasize
that the VMD(4) is derived from this theory, however, the
$\rho\pi\pi$ coupling in this theory is no longer a constant.
There is a function $f_{\rho\pi\pi}(q^2)$ in the vertex ${\cal
L}_{\rho\pi\pi}$. $f_{\rho\pi\pi}(q^2)$
is the intrinsic form factor of pion. The
intrinsic form factor is the physical effect of quark loops. There
are intrinsic form factors for kaons too(see below). In this paper
we apply VMD(4) in which an intrinsic form factor is included to
study form factors of pion and kaons. A
detailed comparison between theoretical results of the form factors
and experimental data in both time-like and space-like regions is presented.

The VMD(4) shows that the pion form factor is determined by two Feynman
diagrams shown in Fig.1 and obtained from Eqs.(4,5,6)
\begin{eqnarray}
|F_\pi (q^2)|^2 & = & f^2_{\rho \pi \pi }(q^2)
\frac{m_\rho ^4+q^2\Gamma _\rho
^2(q^2)}{(q^2-m_\rho ^2)^2+q^2\Gamma _\rho ^2(q^2)}
\end{eqnarray}
in time-like region, where $\Gamma _\rho(q^2)$ is the decay width of $\rho$ 
meson.
\begin{eqnarray}
\Gamma _\rho (q^2) & = & \Gamma _{\rho ^0\rightarrow \pi ^{+}\pi
^{-}}(q^2)+\Gamma _{\rho ^0\rightarrow K\overline{K}}(q^2),
\nonumber \\ \Gamma _{\rho ^0\rightarrow \pi ^{+}\pi ^{-}}(q^2) &
= & \frac{f_{\rho \pi \pi }^2(q^2)\sqrt{q^2}}{12\pi g^2
}(1-\frac{4m_{\pi ^{+}}^2}{q^2})^{{3\over2}}\theta(q^2>4m_{\pi ^{+}}^2),
\nonumber
\\ \Gamma _{\rho ^0\rightarrow K\bar{K}}(q^2) & = & \frac{f_{\rho
\pi \pi }^2(q^2)\sqrt{q^2}}{48\pi g^2
}(1-\frac{4m_{K^{+}}^2}{q^2})^{{3\over2}}\theta(q^2>4m_{K^{+}}^2)
\nonumber \\ &&+\frac{f_{\rho \pi \pi }^2(q^2)\sqrt{q^2}}{48\pi
g^2}(1-\frac{4m_{K^0}^2}{q^2})^{{3\over2}}\theta (q^2>4m_{K^{0}}^2),
\end{eqnarray}
when $q^2>4m_{K^{+}}^2$ the $K\bar{K}$ channel is open. There are
other channels, however, in the range of $\sqrt{q^2}<1.4GeV$ the
contribution of other channels is negligible. 
In
space-like region the pion form factor is
\begin{equation}
F_\pi(q^2)=\frac{m^2_{\rho}f_{\rho\pi\pi}(q^2)}{m^2_\rho-q^2}.
\end{equation}
From Eqs.(7,9) it can be seen that the pion form factor consists
of two parts: the intrinsic form factor $f_{\rho\pi\pi}(q^2)$(6) and
the $\rho-$ pole. The expression of $f_{\rho\pi\pi}(q^2)$ shows that in
time-like region $f_{\rho\pi\pi}(q^2)$ increases with $q^2$ and in
space-like region it decreases with $q^2$. As mention above this
behavior is needed to fit the data.

In time-like region the pion form factor has been measured in
$ee^+ \rightarrow\pi\pi$ and $\tau\rightarrow \pi\pi\nu$. The
cross section of $ee^+ \rightarrow\pi^+ \pi^-$ is expressed as
\begin{eqnarray}
\sigma & = & \frac{\pi \alpha ^2}{3}\frac 1{q^2}(1-\frac{4m_{\pi
^{+}}^2}{q^2})^{{3\over2}}|F_\pi(q^2)|^2.
\end{eqnarray}
Taking $\rho-\omega$ mixing into account 
we obtain 
\begin{eqnarray}
F_\pi(q^2)&=&f_{\rho\pi\pi}(q^2)\{1-\frac{q^2\cos \theta
}{q^2-m_\rho ^2 +i\sqrt{q^2}\Gamma _\rho (q^2)}-\frac
13\frac{q^2\sin \theta }{q^2-m_\omega ^2+i\sqrt{q^2}\Gamma
_\omega }\},
\end{eqnarray}
where $\theta$ is the mixing angle and determined to be
$1.74^0$ in Ref.[6]. From Eq.(11), the pion form factor with $\rho-\omega$
mixing in
space-like region is obtained 
\begin{eqnarray}
F_\pi(q^2)&=&f_{\rho\pi\pi}(q^2)\{1-\frac{q^2\cos \theta
}{q^2-m_\rho ^2}-\frac 13\frac{q^2\sin \theta }{q^2-m_\omega ^2
}\}.
\end{eqnarray}
The comparison of 
the cross section(10), the form factor of pion in both time-like and space-like regions(11-12)
with data are shown in Fig.2-5. Up to
\(\sqrt{q^2}=1.4GeV\) the channel $\rho\rightarrow\pi\pi$ is
dominant and the contribution of the $K\bar{K}$ mode is very
small. 
The intrinsic form factor $f_{\rho\pi\pi}(q^2)$ plays more important role
in the range of higher $q^2$. Using Eq.(8), we
obtain \(\Gamma_{\rho} =143.2MeV\)
which fits the data well(see Fig.2,3). $f^2_{\rho\pi\pi}(q^2)
=1.31$ at \(q^2=m^2_{\rho}\),
 therefore, the intrinsic form factor makes significant contribution
to $\Gamma_\rho$. 
In Ref.[9] a fitting of $\rho$
resonance to $ee^+\rightarrow\pi\pi$ has been presented. The
radius of charged pion is found from Eq.(12)
\begin{eqnarray}
<r^2>_\pi & = & 6(\frac{\cos \theta }{m_\rho ^2}+\frac{\sin \theta
}{3m_\omega ^2})+\frac 3{\pi ^2f_\pi ^2}\{(1-\frac{2c}g)^2-4\pi
^2c^2\}.
\end{eqnarray}
The numerical result is
\begin{equation}
<r^2>_\pi =(0.395+0.057)fm^2 = 0.452fm^2,
\end{equation}
where the first number 0.395 comes from $\rho$ and $\omega$ poles
and the second is the contribution of the intrinsic form
factor(6), which is about $13\%$ of the total value. The
experimental data is $(0.439\pm 0.03)fm^2$ [3].

In terms of the method presented in Ref.[7], the decay rate of
$\tau\rightarrow\pi\pi\nu$ is derived
\begin{eqnarray}
\frac{d\Gamma }{dq^2}&=&\frac{G^2}{(2\pi )^3}\frac{\cos ^2\theta
_C}{48m_\tau ^3}(m_\tau ^2+2q^2)(m_\tau ^2-q^2)^2\{1-\frac{4m_\pi
^2}{q^2}\}^{3/2}|F_\pi (q^2)|^2,
\end{eqnarray}
where $F_\pi(q^2)$ is given by Eq.(7). In Ref.[10] it is shown that the
form factor $F_\pi(q^2)$ determined from $\tau$ decay is
consistent with the one obtained from $ee^+$ annihilation. The
branching ratio is calculated to be
\begin{eqnarray}
B_{\tau ^{-}\rightarrow
\pi ^0\pi ^{-}\nu _\tau }&=&22.3\%.
\end{eqnarray}
The experimental data is $B _{\tau ^{-}\rightarrow \pi ^0\pi
^{-}\nu _\tau }=(25.32\pm 0.15)\%$[11].

In the same way the form factors of charged and neutral kaons are
studied. All the three vector mesons, $\rho, \omega$, and $\phi$,
contribute to the form factors of kaons. Using the substitutions
\begin{equation}
\phi _\mu \longrightarrow -\frac{eg}{3\sqrt{2}}A_\mu,
\;\;\; \omega _\mu \longrightarrow
\frac{eg}6A_\mu,
\end{equation}
the currents $j_\mu^{\omega,\phi}$ of Eq.(4) are obtained from the vertices[6]
\begin{eqnarray}
{\cal L}_{vK\overline{K}}&=&-\frac{2\sqrt{2}}{g}if_{\rho \pi \pi
}(q^2)\phi ^\mu (K^{+}\partial _\mu K^{-}+K^0\partial _\mu
\overline{K^0}) \nonumber \\ &&+\frac{2}{g}if_{\rho \pi \pi
}(q^2)\omega ^\mu (K^{+}\partial _\mu K^{-}+K^0\partial _\mu
\overline{K^0}) \nonumber \\ &&+\frac{2}{g}if_{\rho \pi \pi }(q^2)\rho
^\mu (K^{+}\partial _\mu K^{-}-K^0\partial _\mu \overline{K^0}).
\end{eqnarray}
The cross sections of e$^{+}$e$^{-}\longrightarrow K^{+} K^{-}$
and e$^{+}$e$^{-}\longrightarrow K^0\overline{K^0}$ are derived
\begin{eqnarray}
\sigma _{e^{+}e^{-}\rightarrow K^{+}K^{-}}&=&\frac{\pi \alpha
^2}3\frac 1{q^2}(1-\frac{4m_{K^{+}}^2}{q^2})^{{3\over2}}
|F_{K^+}|^2, \nonumber \\
\sigma _{e^{+}e^{-}\rightarrow K^0\overline{K^0}}&=&\frac{\pi
\alpha ^2}3\frac 1{q^2}(1-\frac{4m_{K^{0}}^2}{q^2})^{{3\over2}}
|F_{K^0}|^2,
\end{eqnarray}
where
\begin{eqnarray}
|F_{K^{+}}(q^2)|^2&=&f^2_{\rho \pi \pi }(q^2)|A|^2,\\
|F_{K^{0}}(q^2)|^2&=&f^2_{\rho \pi \pi }(q^2)|B|^2.
\end{eqnarray}
A and B are defined as
\begin{eqnarray}
A&=&\frac 12\frac{
-m^2_\rho+i\sqrt{q^2}\Gamma_\rho(q^2)}
{q^2-m_\rho^2+i\sqrt{q^2}\Gamma _\rho
(q^2)}+\frac 16
\frac{-m^2_\omega+i\sqrt{q^2}\Gamma_\omega}
{q^2-m_\omega ^2+i\sqrt{q^2}\Gamma
_\omega }, \nonumber
\\&&+\frac 13
\frac{-m^2_\phi+i\sqrt{q^2}\Gamma_\phi(q^2)}
{q^2-m_\phi ^2+i\sqrt{q^2}\Gamma
_\phi(q^2) },\\ 
B&=&-\frac 12\frac{
-m^2_\rho+i\sqrt{q^2}\Gamma_\rho(q^2)}
{q^2-m_\rho^2+i\sqrt{q^2}\Gamma _\rho
(q^2)}+\frac 16
\frac{-m^2_\omega+i\sqrt{q^2}\Gamma_\omega}
{q^2-m_\omega ^2+i\sqrt{q^2}\Gamma
_\omega }, \nonumber
\\&&+\frac 13
\frac{-m^2_\phi+i\sqrt{q^2}\Gamma_\phi(q^2)}
{q^2-m_\phi ^2+i\sqrt{q^2}\Gamma
_\phi(q^2) } 
\end{eqnarray}
with
\begin{eqnarray}
\Gamma _\phi (q^2)&=&\Gamma _{\phi \rightarrow K^{+}K^{-}}(q^2)
+\Gamma _{\phi \rightarrow K^0\overline{K^0}}(q^2), \nonumber
\\
\Gamma _{\phi \rightarrow
K^{+}K^{-}}(q^2)&=&\frac{\sqrt{q^2}}{24g^2 \pi }f_{\rho \pi \pi
}^2(q^2)(1-\frac{4m_{K^{+}}^2}{q^2})^{{3\over2}}, \nonumber
\\
\Gamma _{\phi \rightarrow
K^0\overline{K^0}}(q^2)&=&\frac{\sqrt{q^2}}{24g^2 \pi }f_{\rho \pi
\pi }^2(q^2)(1-\frac{4m_{K^0}^2}{q^2})^{{3\over2}}.
\end{eqnarray}
The numerical results are
\begin{eqnarray}
\Gamma(\phi \rightarrow K^+ K^-)&=&2.14MeV,\;\;\; \Gamma (\phi
\rightarrow K^0 \overline{K}^0)=1.4MeV,\nonumber \\ \Gamma(\phi
\rightarrow K^+ K^-) _{exp}&=&2.18(1\pm 0.028)MeV, \nonumber \\
\Gamma(\phi \rightarrow K^0 \bar{K}^0) _{exp} &=&1.51(1\pm
0.029)MeV.
\end{eqnarray}
$f^2_{\rho\pi\pi}(q^2)=1.57$ at $q^2=m^2_\phi$. 
The contribution of the intrinsic forn factor is significant. 
Theoretical results
of $\Gamma_\phi$ agree with the data very well.
$\Gamma_\omega$ has been calculated to be 7.7MeV[7] which is
consistent with the data $7.49(1\pm0.02)$MeV.

The cross sections of $ee^+\rightarrow K\bar{K}$ and the form
factors of kaons in time-like region are shown in Fig.6-9.

The form factors of kaons in space-like region are obtained from
Eqs.(20-23)
\begin{eqnarray}
F_{K^{+}}(q^2)&=&f_{\rho \pi \pi }(q^2) \{\frac
12\frac{m_\rho ^2}{m_\rho ^2-q^2}+\frac
16\frac{m_\omega ^2}{m_\omega ^2-q^2}+\frac 13\frac{m_\phi
^2}{m_\phi ^2-q^2}\}, \nonumber \\ F_{K^0}(q^2)&=&f_{\rho
\pi \pi }(q^2) \{-\frac{1}2\frac{m_\rho ^2}{ m_\rho ^2-q^2}
+\frac 16\frac{m_\omega ^2}{m_\omega
^2-q^2}+\frac 13\frac{m_\phi ^2}{m_\phi ^2-q^2}\}.
\end{eqnarray}
The form factor, $F_{K^{+}}(q^2)$, is shown in Fig.10. The radius of
the charged kaon is derived from Eq.(26)
\begin{eqnarray}
<r^2>_{K^{+}}&=&\frac 3{\pi ^2f_\pi ^2}\{(1-\frac{2c}g)^2-4\pi
^2c^2\}+(\frac 3{m_\rho ^2}+\frac 2{m_\phi ^2}+\frac 1{m_\omega
^2}).
\end{eqnarray}
The numerical value is
\begin{equation}
<r^2>_{K^{+}}=(0.05+0.33)fm^2
=0.38fm^2.
\end{equation}
The first number comes from the intrinsic form factor. The
experimental data is $(0.34\pm 0.05)fm^2$ [16]. The radius of the
neutral kaon is obtained from Eq.(26)
\begin{eqnarray}
<r^2>_{K^0}=-6\frac{\partial F_{K^0}(q^2)}{\partial q^2}\mid _{q^2=0}
=-\frac{2}{m_{\phi ^2}}-\frac 1{m_\omega ^2}+\frac
3{m_\rho ^2}=0.057fm^2.
\end{eqnarray}

Only the coupling of $\rho K\bar{K}$ in Eq.(18) contributes to
$\tau^-\rightarrow K^0 K^-\nu$. The decay rate is found to be
\begin{eqnarray}
\frac{d\Gamma }{dq^2}&=&\frac{G^2}{(2\pi )^3}\frac{\cos ^2\theta
_C}{96m_\tau ^3}(m_\tau ^2+2q^2)(m_\tau
^2-q^2)^2\{1-\frac{4m_K
^2}{q^2}\}^{3/2}|F_\pi (q^2)|^2,
\end{eqnarray}
where $F_\pi(q^2)$ is given by Eq.(7). The theory predicts that 
the decay rate of
this process is determined by the pion form factor in the chiral limit.
So far there is no measurement on this distribution. The branching ratio of
this decay mode is calculated to be
\begin{eqnarray}
B_{\tau ^{-}\longrightarrow K^0K^{-}\nu _\tau }&=&1.78\times
10^{-3}.
\end{eqnarray}
The experimental data is $(1.59\pm 0.24)\times 10^{-3}$[11].
Theory agrees with the data within the error bar.

In summary, the VMD 
derived from the chiral theory[6] has been applied to study the form factors 
of pion and kaons. An intrinsic form factor has been predicted by this 
chiral theory[6]. It has been found that this intrinsic form factor makes 
significant contribution to the form factors of pion and kaons, 
the decay widths of $\rho$ and $\phi$ mesons, and the decays 
$\tau\rightarrow\pi\pi\nu$ and $\tau\rightarrow K\bar{K}\nu$. 
The theory agrees well with the data.
It is necessary
to point out that there is no new parameter in this study. 
In this theory derivative expansion is used and all the calculations are done up to the fourth
order in derivatives. The fitting shows that the pion form factor is in 
agreement with data up to $q\sim1.4GeV$ in time-like region(Fig.2,3). 
In the range of higher $q^2$ the contribution of higher order derivatives should be taken into account. We will provide the study in the
near future.    

The study is supported by DOE grant No.DE-91ER75661.

\pagebreak

\pagebreak
\begin{flushleft}
{\bf Figure Captions}
\end{flushleft}
{\bf FIG. 1.} Feynman Diagrams of pion form factor.
\\ {\bf FIG. 2.}
Cross Section of e$^{+}e^{-}->\pi ^{+}\pi ^{-}$ vs invariant mass
of $\pi \pi $. Data are from Ref.[12].
\\{\bf FIG. 3.} Pion form factor in time-like region. Data 
are from [12].
\\ {\bf FIG. 4.} Pion form factor in space-like region.
Data are from [13].
\\{\bf FIG. 5.} Pion form factor in space-like region.
\\{\bf FIG. 6.} Cross Section of e$^{+}e^{-}->K^{+}K^{-}$ vs
invariant mass of KK. Data are from [14,15].
\\{\bf FIG. 7.} Cross Section of e$^{+}e^{-}->K^0\overline{K^0}$
vs invariant mass of KK. Data are from [14,15].
\\{\bf FIG. 8.} Charged kaon form factor in time-like region. Data
are from [14-16].
\\{\bf FIG. 9.} Neutral kaon form factor in time-like region. Data
are from [14-16].
\\{\bf FIG. 10.} Charged kaon form factor in space-like region.

\begin{figure}
\begin{center}
\includegraphics[width=7in, height=7in]{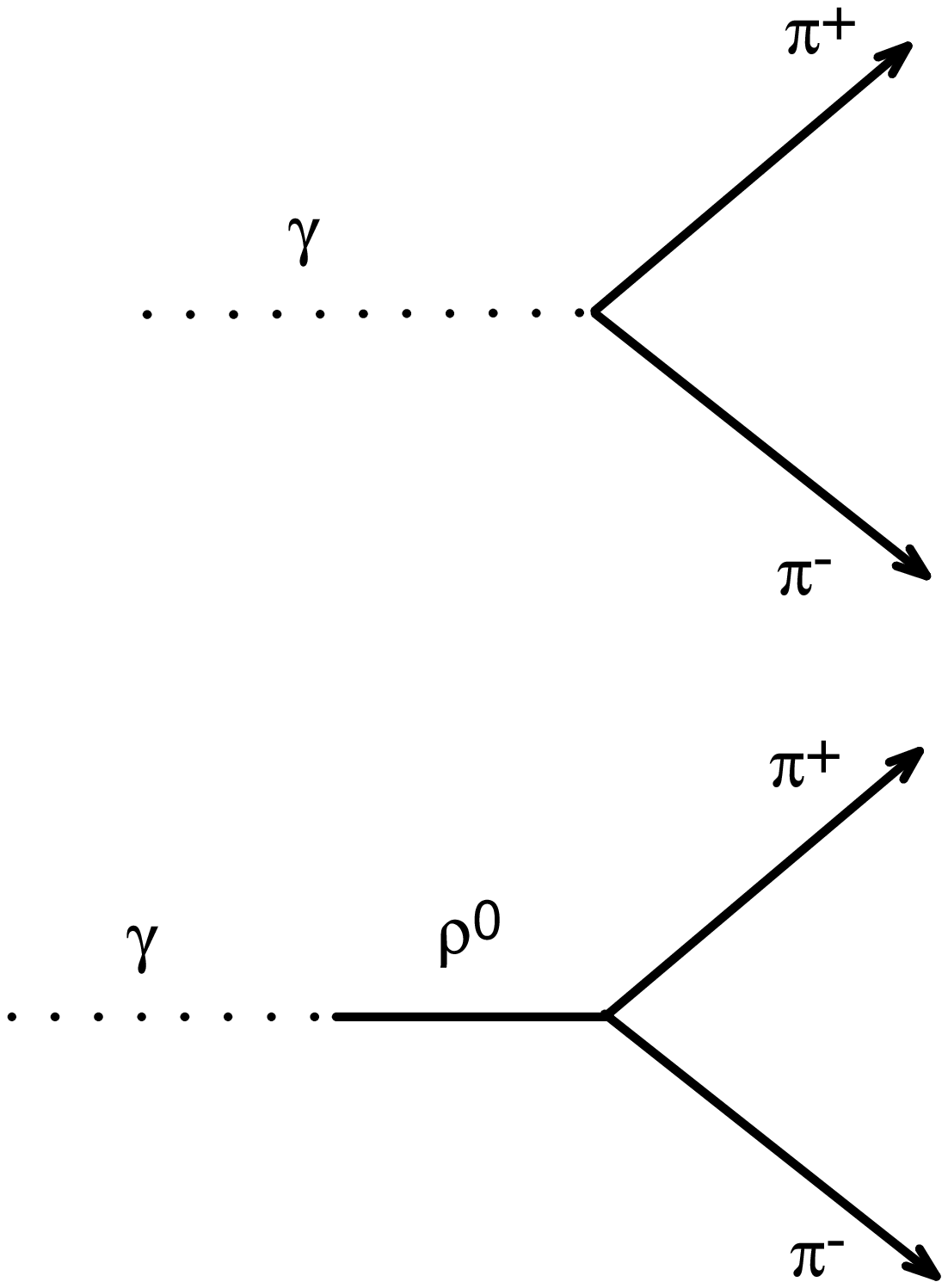}
FIG. 1.
\end{center}
\end{figure}

\begin{figure}
\begin{center}
\includegraphics[width=7in, height=7in]{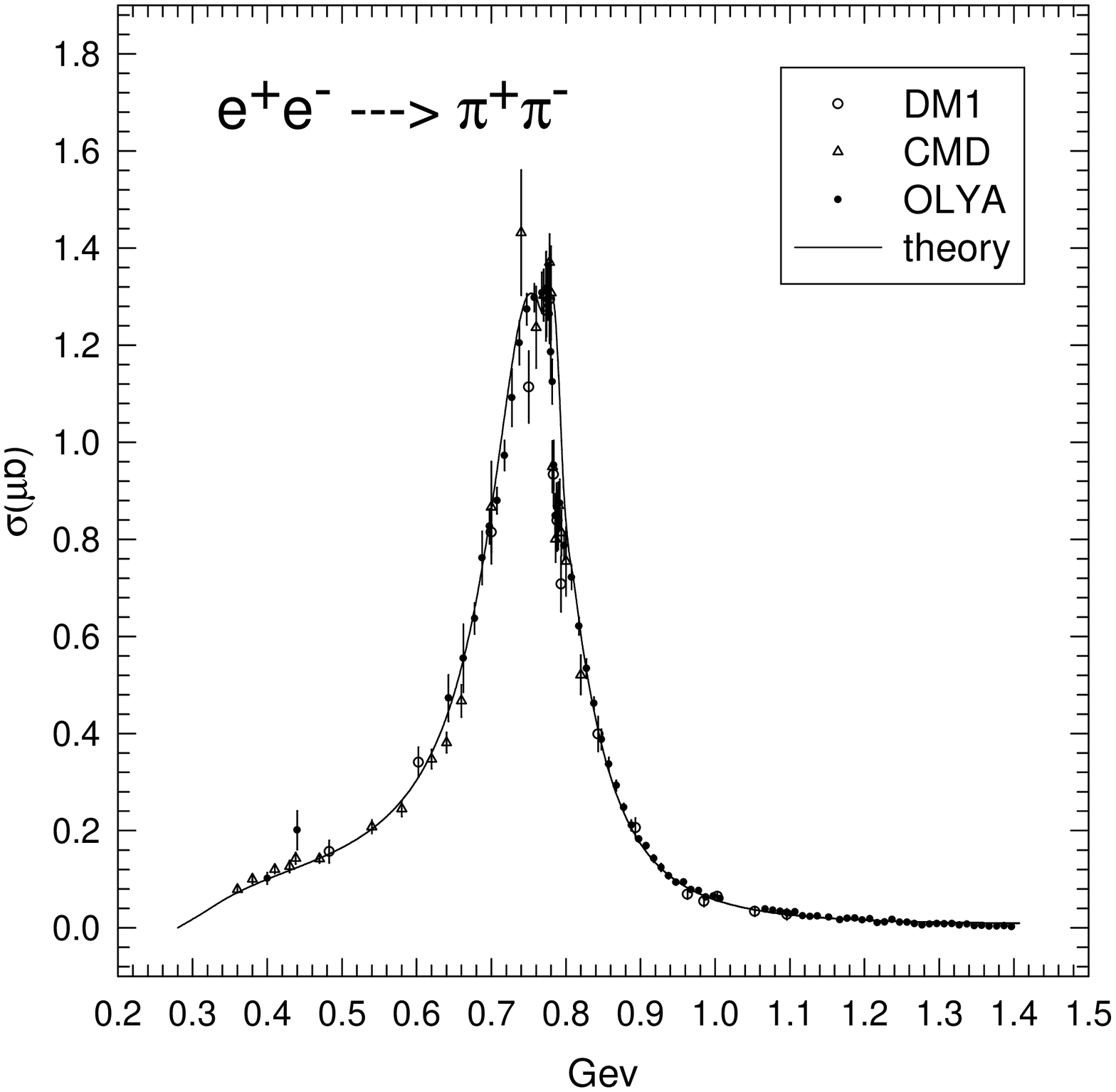}
FIG. 2.
\end{center}
\end{figure}

\begin{figure}
\begin{center}
\includegraphics[width=7in, height=7in]{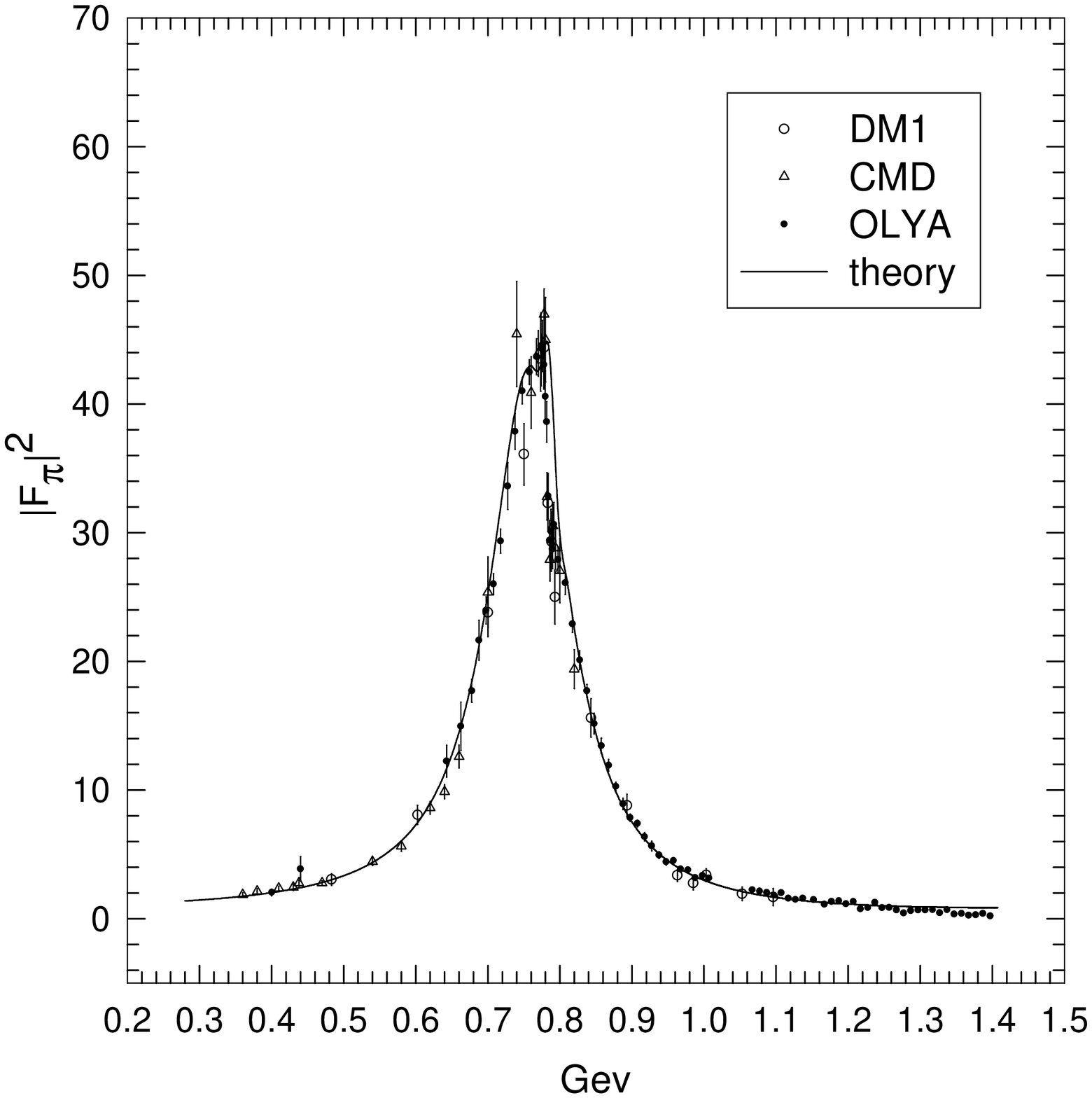}
FIG. 3.
\end{center}
\end{figure}

\begin{figure}
\begin{center}
\includegraphics[width=7in, height=7in]{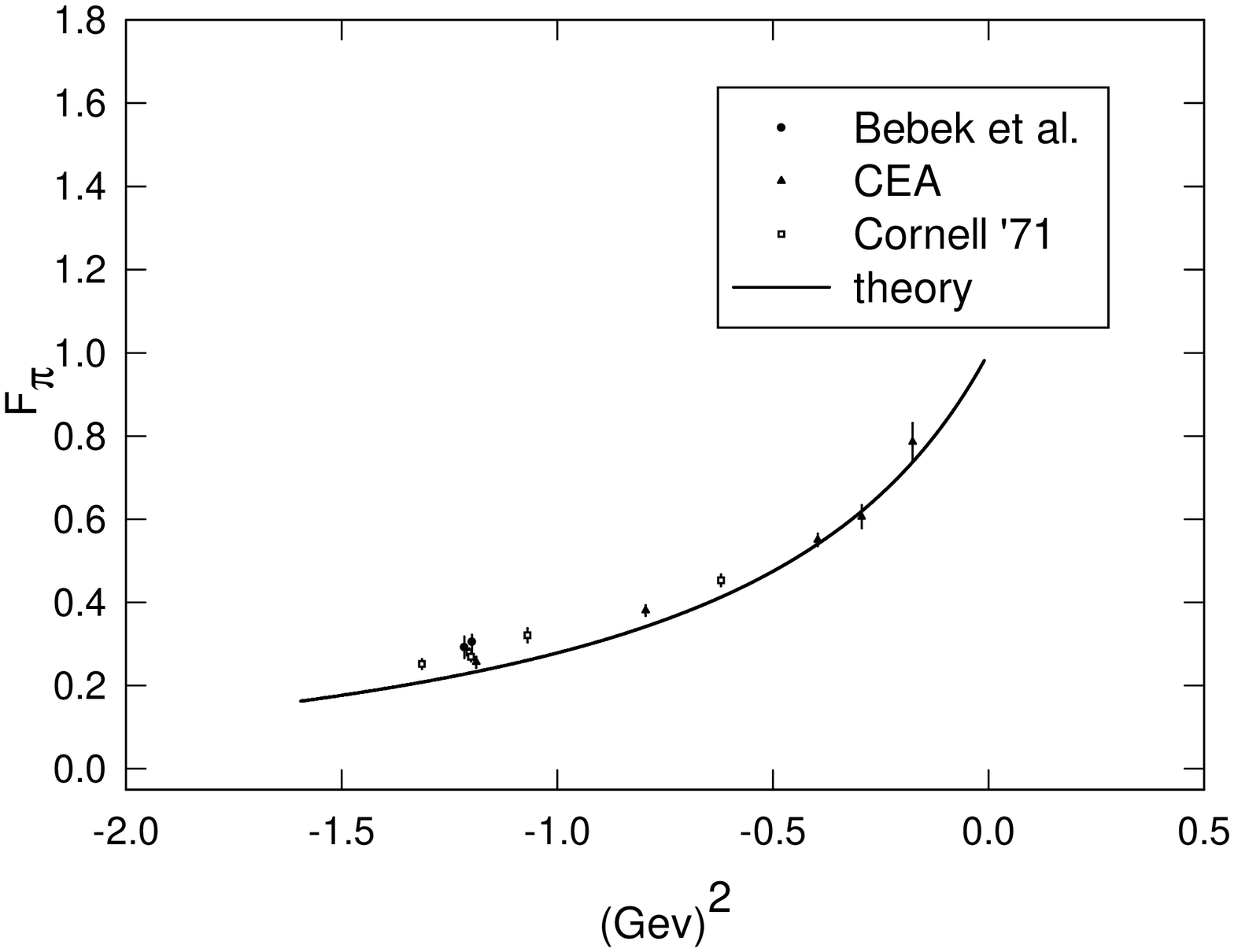}
FIG. 4.
\end{center}
\end{figure}

\begin{figure}
\begin{center}
\includegraphics[width=7in, height=7in]{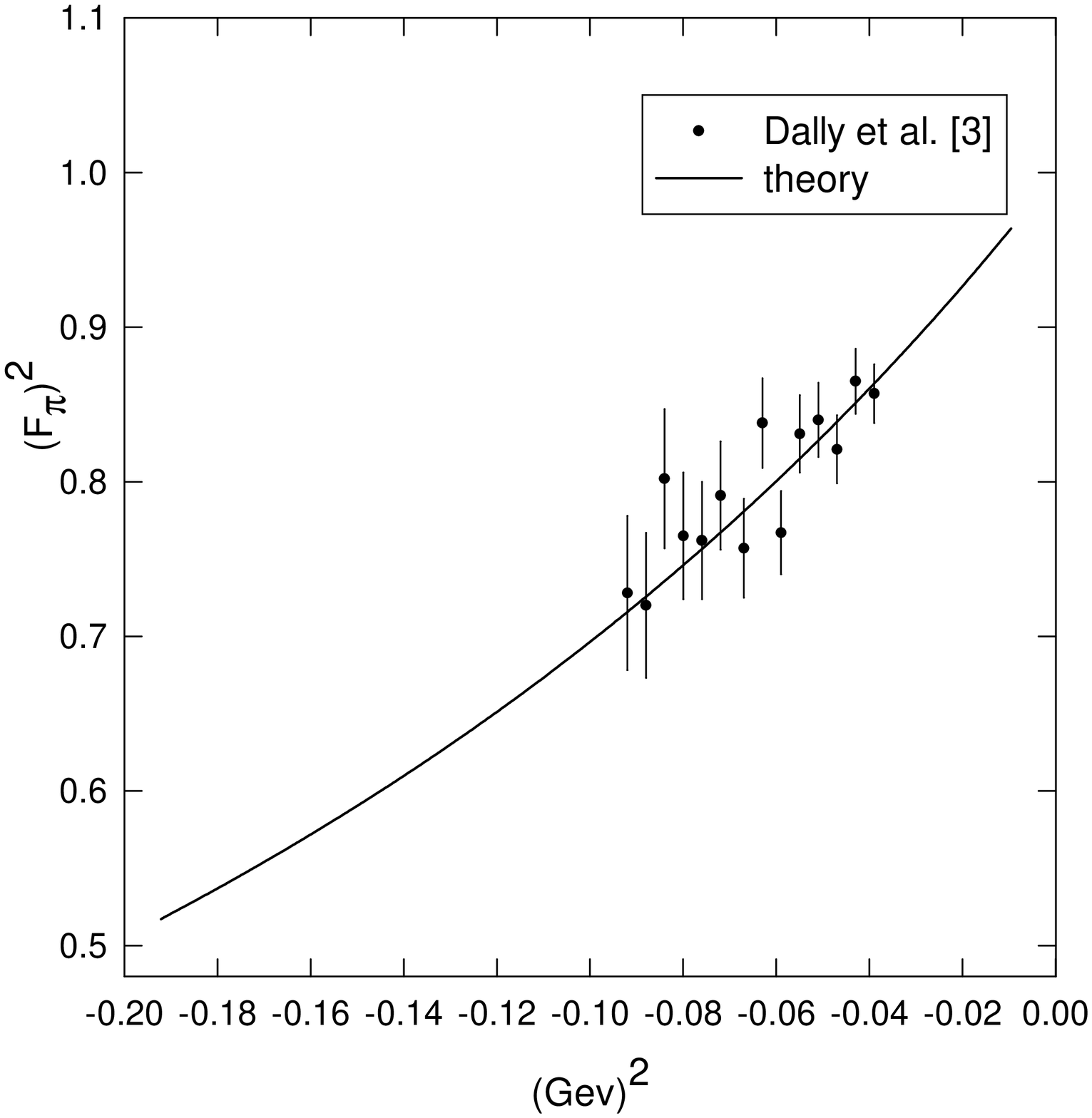}
FIG. 5.
\end{center}
\end{figure}

\begin{figure}
\begin{center}
\includegraphics[width=7in, height=7in]{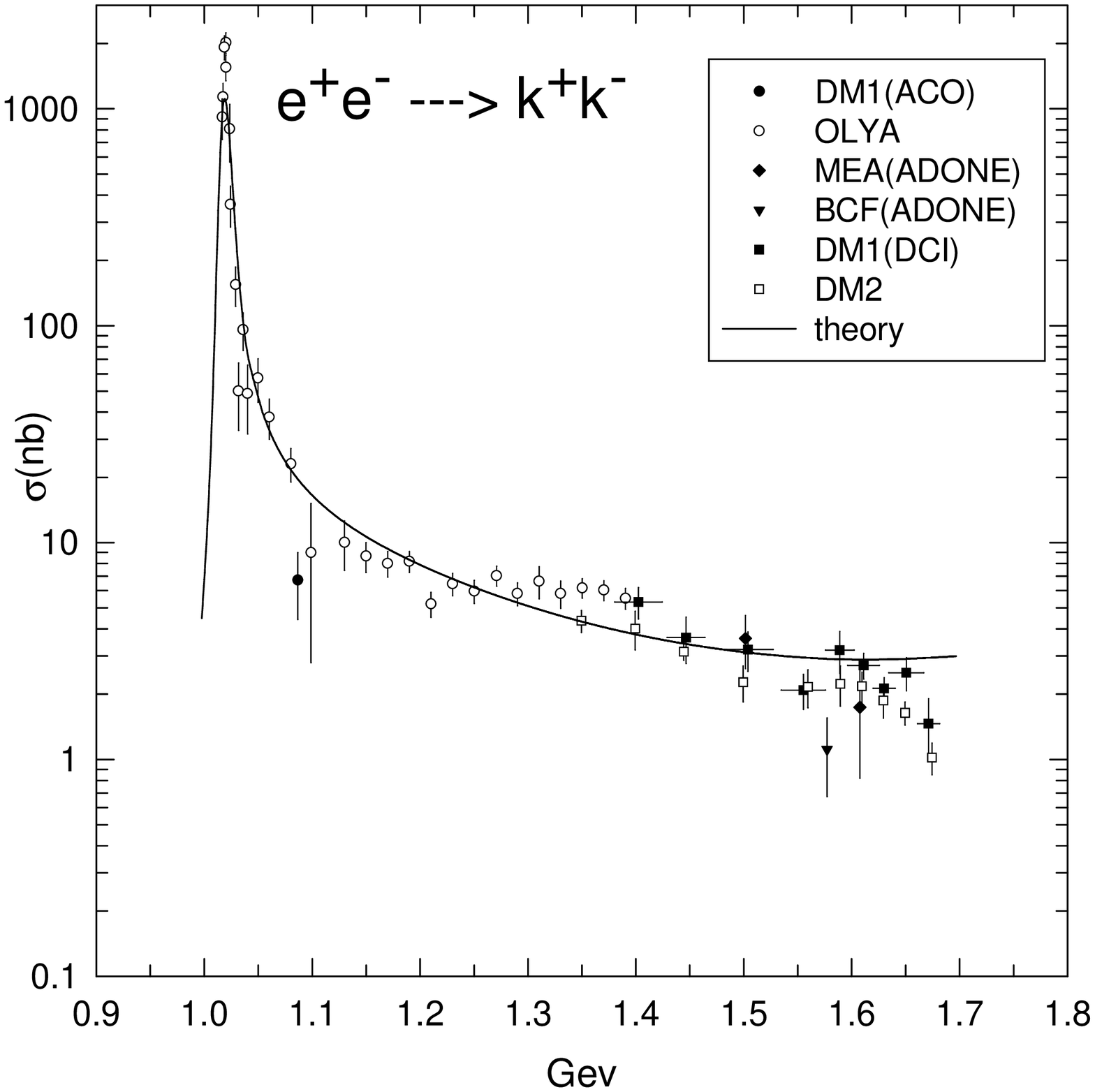}
FIG. 6.
\end{center}
\end{figure}

\begin{figure}
\begin{center}
\includegraphics[width=7in, height=7in]{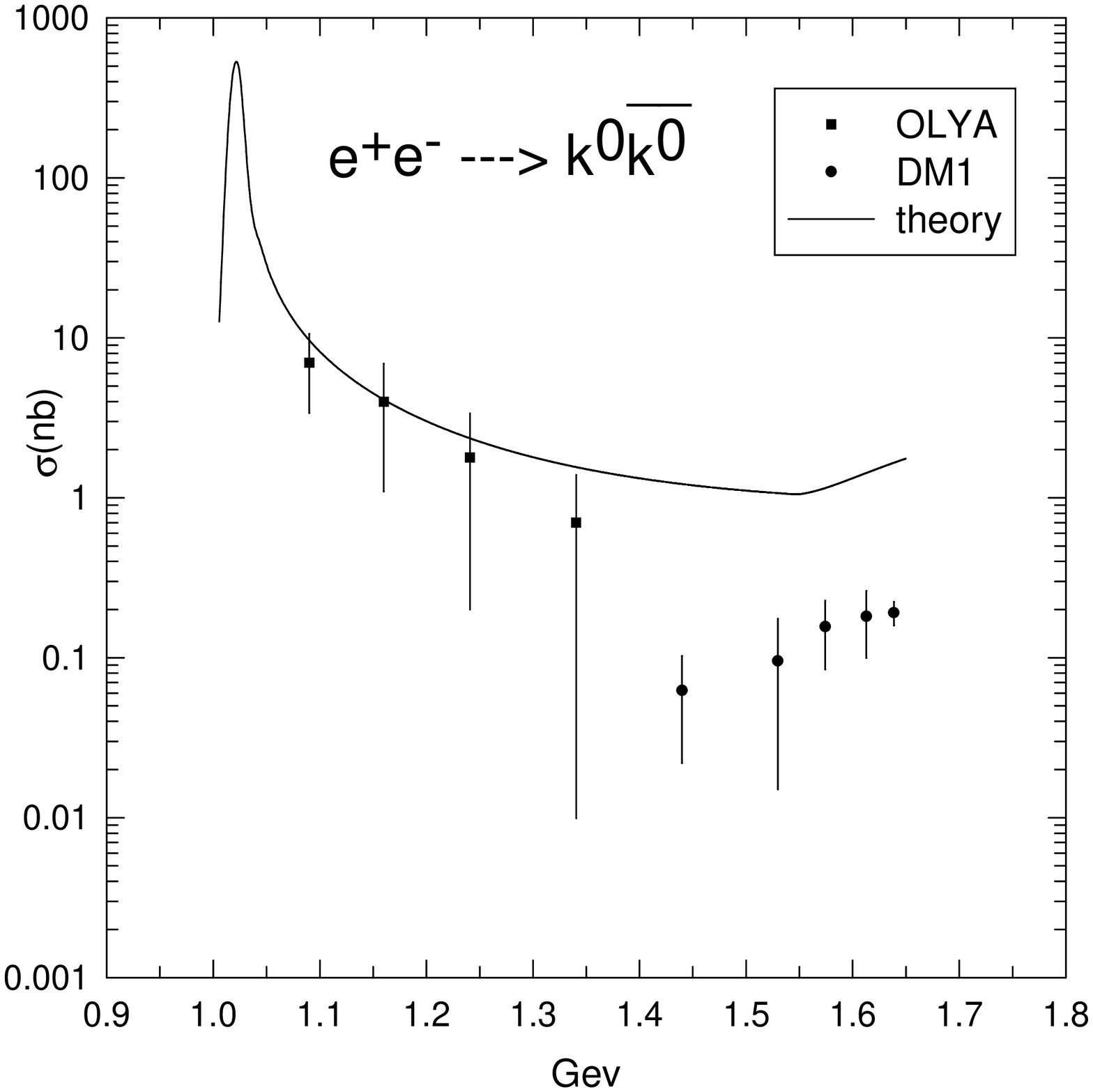}
FIG. 7.
\end{center}
\end{figure}

\begin{figure}
\begin{center}
\includegraphics[width=7in, height=7in]{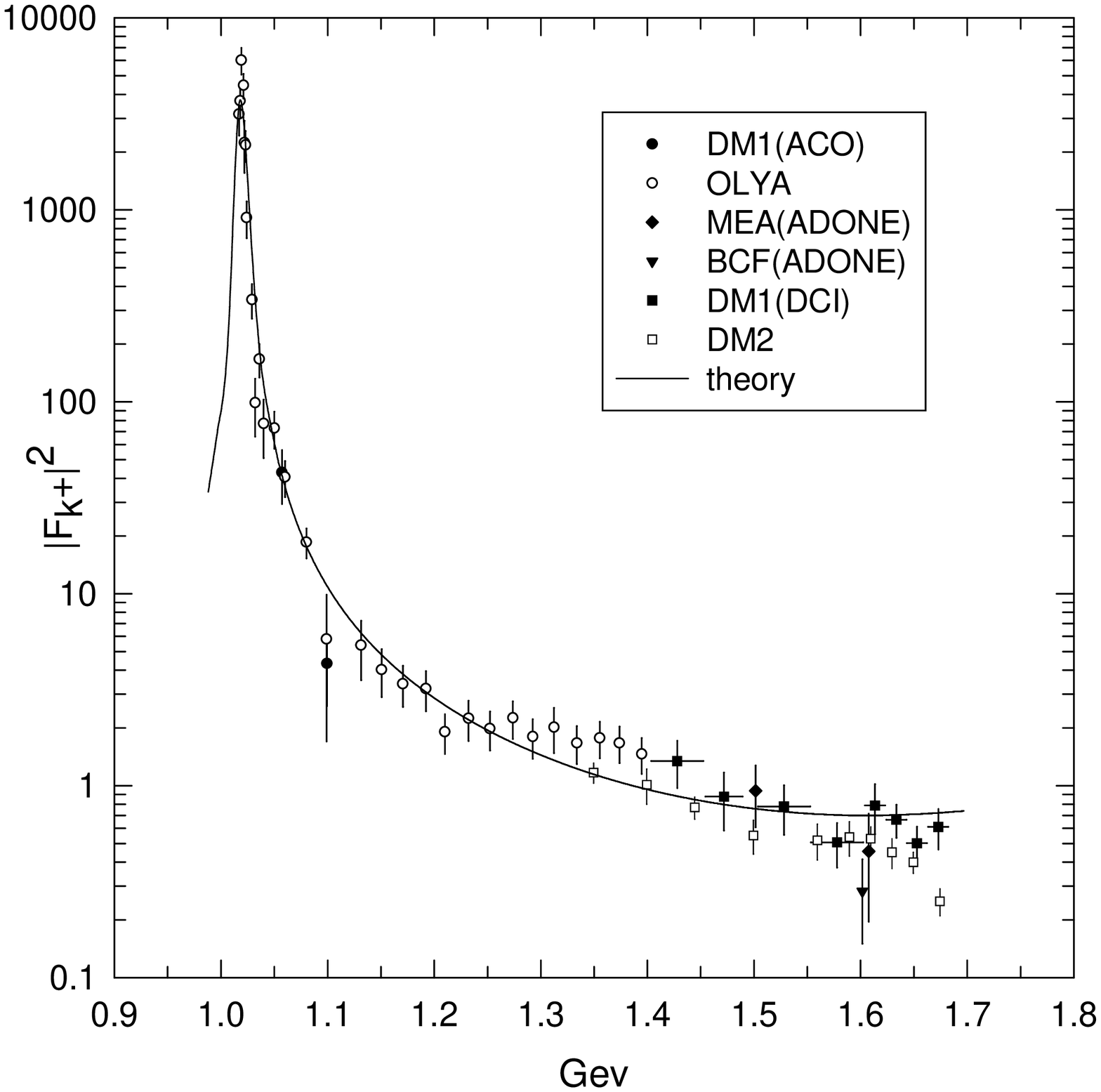}
FIG. 8.
\end{center}
\end{figure}

\begin{figure}
\begin{center}
\includegraphics[width=7in, height=7in]{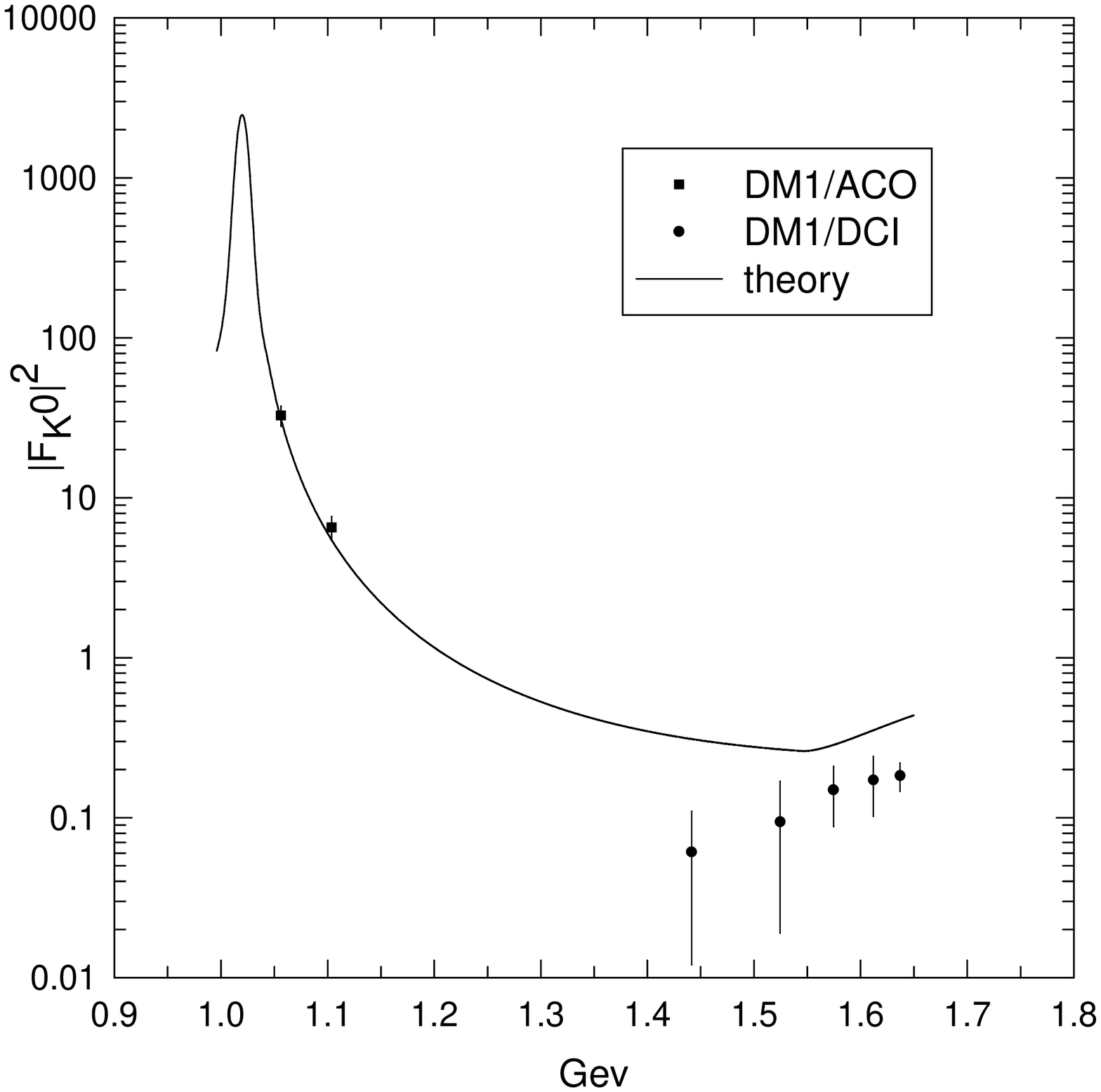}
FIG. 9.
\end{center}
\end{figure}

\begin{figure}
\begin{center}
\includegraphics[width=7in, height=7in]{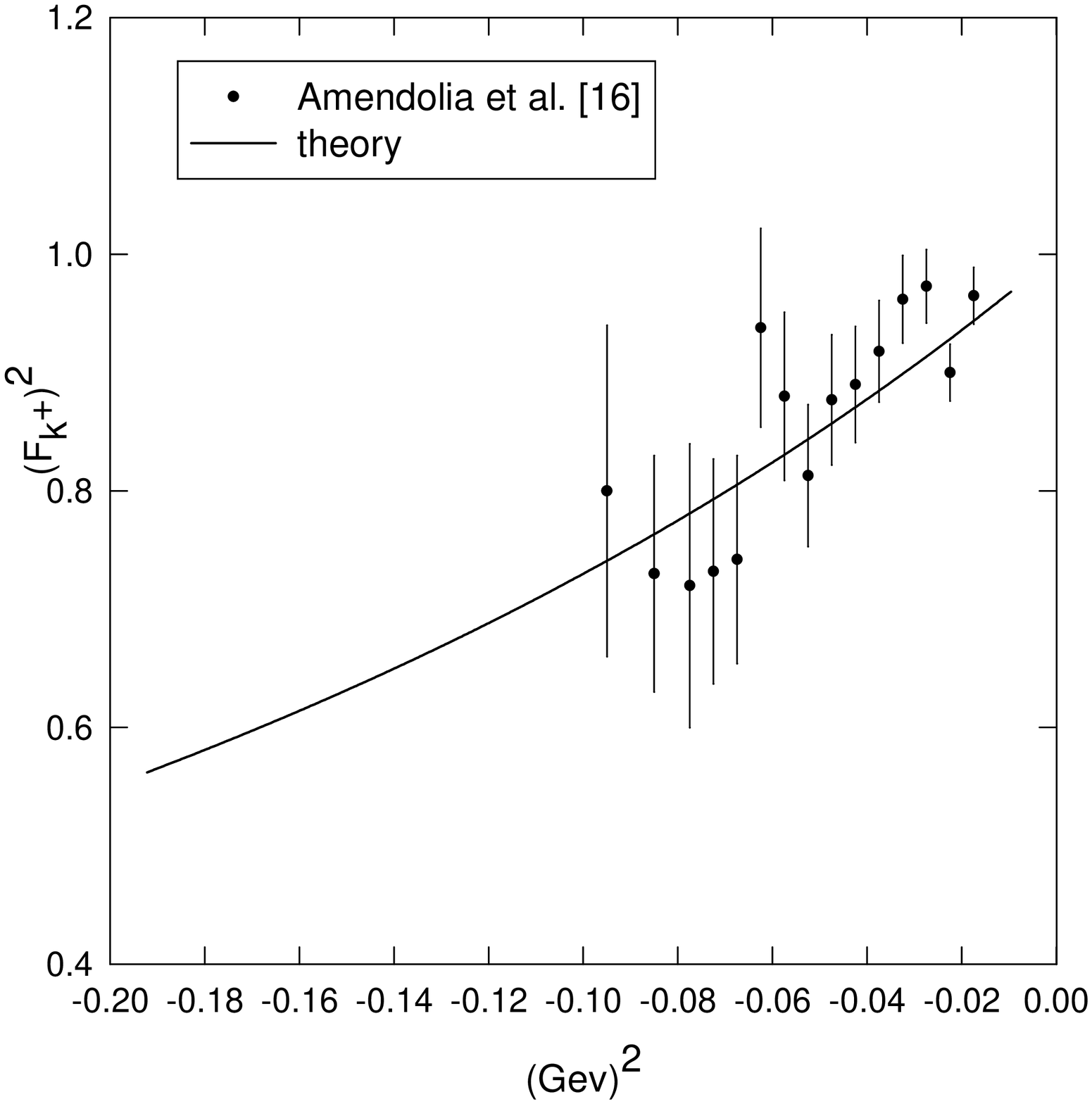}
FIG. 10.
\end{center}
\end{figure}

\end{document}